\documentstyle[12pt,epsfig]{aipproc}
\def\rf#1#2#3{{\bf #1}, #2 (19#3)}
\def\rft#1#2#3{{\bf #1}, #2 (20#3)}
\def\np{Nucl.\ Phys.\ }
\def\pl{Phys.\ Lett.\ }
\def\pr{Phys.\ Rev.\ }
\def\prl{Phys.\ Rev.\ Lett.\ }
\def\ppnp{Prog.\ Part.\ Nucl.\ Phys.\ }

\begin{document}

\title{The Shape and Experimental Tests of the\\ $Q^2$-Invariant Polarized
Gluon Asymmetry}
\author
{Gordon P. Ramsey\\Physics Dept., Loyola University Chicago and\\
HEP Division, Argonne National Lab
\footnote{Talk given at the Spin Physics Symposium (SPIN 2000),
16-21 October 2000, Osaka, Japan. Based upon work done with F. Close and D.
Sivers.}$^,$\footnote{Work supported by the U.S. Department of Energy,
Division of High Energy Physics, Contract W-31-109-ENG-38. E-mail:
gpr@hep.anl.gov}}
\maketitle

\begin{abstract}
The absence of "valence-gluon`` degrees of freedom combined with an examination
of radiative QCD diagrams leads to an implication that the gluon spin asymmetry
in a proton, defined as $A_G(x,Q^2)={{\Delta G(x,Q^2)}\over {G(x,Q^2)}}$,
should be approximately $Q^2$ invariant. The condition for scale invariance
completely determines the $x$-dependence of this asymmetry, which satisfies
constituent counting rules and reproduces the basic results of the 
Bremsstrahlung model originated by Close and Sivers. This asymmetry can be
combined with the measured unpolarized gluon density, $G(x,Q^2)$ to provide a
prediction for $\Delta G(x,Q^2)$. Existing and proposed experiments can test
both the prediction of scale-invariance for $A_G(x,Q^2)$ and the nature of
$\Delta G$ itself.
\end{abstract}

\section{Introduction}

The spin-weighted gluon density, $\Delta G(x,Q^2)$, is of fundamental
importance in understanding the dynamics of hadron structure. Numerous
experiments have been proposed\cite{hermes,rhic,compass} to determine this 
distribution experimentally. 
Measurements of the deep-inelastic scattering asymmetry $A_1(x,Q^2)$
for protons, neutrons and deuterons yield data from which polarized quark
distributions may be inferred, but the shape and size of the polarized gluon
density has not been determined. However, the constituent quark model provides
a framework for predicting an essential feature of $\Delta G(x,Q^2)$. To
understand this, we assume that the spin structure of proton does not have a
significant component representing a valence or ``constituent" gluon
polarization. Hadronic spin observables at
small $Q^2$ conform to the non-relativistic quark model in which spin degrees
of freedom are associated with constituent quarks. This assumption does
\underline{not} imply that $\Delta G\to 0$ at low $Q^2$. In fact, when the
spin structure of the constituent quarks is resolved by inelastic scattering
measurements, this approach yields a variation of Close-Sivers Bremsstrahlung
model\cite{cs} which displays a maximal gluon polarization at $x=1$.

In a positive helicity proton, we define the gluon polarization asymmetry as
\begin{equation} 
A_G(x,t)\equiv \Delta G(x,t)/G(x,t), \label{gpr:1.1}
\end{equation}
where the evolution variable $t\equiv \ln[\alpha_s(Q_0^2)/\alpha_s(Q^2)]$.
It is assumed that the same factorization prescription is used to define all
of the densities in equation (\ref{gpr:1.1}). Since there are no overwhelming
theoretical arguments favoring any single model for $\Delta G$, we consider a
more direct argument for its shape in terms of this asymmetry.

Our results follow from the observation that, in the absence of a
``constituent" gluon, both $G(x,t)$ and $\Delta G(x,t)$ exhibit scaling
violations which can be associated with measurements ``resolving" radiative
diagrams. The diagrams leading to positive and negative helicity gluons are the
same. This implies that the relative probability of measuring a gluon of 
either helicity does not depend upon $t$. Thus, the gluon polarization 
asymmetry is predicted to be scale invariant: $\partial A_G(x,t)/\partial t=0$.
This would not be true if there were a valence gluon, since the shape of $A_G$
would then depend upon the relative amount of valence and radiated gluons. 
It is reasonable to choose $t=0$ to coincide with a typical hadronic scale,
$Q^2=m_H^2$. The scale-invariance assumption provides the $x$ dependence of
$A_G(x)$, which satisfies several important physical constraints:
\begin{itemize}
\item it obeys the constituent-counting rules,
\item for large $x$, where quark distributions dominate the gluon distribution,
the predicted asymmetry coincides with the original QCD-Bremsstrahlung model
of Close and Sivers\cite{cs}. At other values of $x$, it corresponds to a
natural extension of the QCD-Bremsstrahlung approach by allowing for radiation
from both quarks and gluons, and
\item for small $x$, where the gluon distribution is expected to dominate the
quark distributions, the scale-invariant asymmetry arises as a natural
asymptotic limit, independent of the starting point. 
\end{itemize}

Thus, the arguments originally presented in ref. \cite{cs} can be combined with 
existing parametrizations of polarized and unpolarized quark distributions to
provide a quantitative estimate for $\Delta G(x,Q_0^2)$ at any convenient 
reference scale.

\section{The Shape of $A_G(x)$}

In 1977, Close and Sivers\cite{cs} proposed that the quark {\it sea} should be
polarized and that {\it gluons} should exhibit a polarization of in the same 
direction as the proton. This was based on perturbative QCD and the theoretical
understanding that valence quarks are polarized in the same sense as the 
proton. This happens
since the $\gamma_{\mu}$ coupling of the quark-gluon vertex conserves quark
helicity when quark masses are neglected. Thus, when a gluon is radiated by a
quark, its helicity has the same sign as the creating quark.

A phenomenological picture of the proton assumes that at low $Q^2\le m_P^2$, a
proton consists of three ``valence" quarks, surrounded by radiated gluons and
$q\bar q$ pairs. The Bremsstrahlung mechanism used in ref. \cite{cs} supplies
a significant fraction of gluons in a proton found at low to medium values of
$Q^2$. From a reference scale where the constituent
quark picture is applicable, the QCD evolution equations can be used to 
generate a prediction for the quark and gluon distributions at higher $Q^2$.

The requirement that $A_G(x,t)$ has no $t$-dependence implies that
\begin{equation}
{{\partial A_G}\over {\partial t}}={1\over G}\Bigl[{{\partial \Delta G}
\over {\partial t}}-A_G(x,t) {{\partial G}\over {\partial t}}\Bigr]=0. 
\label{gpr:2.1}
\end{equation}
The $t$-dependence of the gluon distributions is given by the corresponding
DGLAP evolution equations.\cite{dglap} Combining the DGLAP equations with 
equation (\ref{gpr:2.1}) gives
\begin{equation}
A_G={{{\partial \Delta G}\over {\partial t}}\over {{\partial G}\over
{\partial t}}}
=\Biggl[{{\Delta P_{Gq} \otimes \Delta q+\Delta P_{GG}\otimes \Delta G}\over
{P_{Gq} \otimes q+P_{GG}\otimes G}}\Biggr]. \label{gpr:2.2}
\end{equation}
This follows since the diagrams which determine the $t$ evolution of the 
distributions are the same as those which distribute the spin information to 
the gluons.

Since $\Delta G$ has not been measured, equation (\ref{gpr:2.2}) can be 
converted into a non-linear equation for $A_G(x)$ by inserting 
$\Delta G(x,t)=A_G(x)\cdot G(x,t)$ into the convolution,
\begin{equation}
A_G=\Biggl[{{\Delta P_{Gq} \otimes \Delta q+\Delta P_{GG}\otimes (A_G\cdot G)}
\over {P_{Gq} \otimes q+P_{GG}\otimes G}}\Biggr]. \label{gpr:2.3}
\end{equation}

An equation in this form can be solved iteratively. We first observe that for
a given value of $x$, the distributions in the DGLAP equations enter only in
the range $[x,1]$. Then, for a large enough $x$ ($x\geq 0.6$), the gluon
distributions on the right side of (\ref{gpr:2.3}) can be neglected. Now, the
polarized DIS data are consistent with the constituent counting rule result that
$\lim_{x\to 1}\> A_1(x,Q^2)\approx \lim_{x\to 1}\>\Delta u_v(x,Q^2)/u_v(x,Q^2)
=1.$ There exist parametrizations of the helicity-weighted quark
distributions \cite{ggr} which incorporate this result to reproduce all of the
existing data. Thus, we make an initial approximation
\begin{equation}
\lim_{x\to 1}A_G^0=\Biggl[{{\Delta P_{Gq} \otimes \Delta u_v} \over {P_{Gq}
\otimes u_v}}\Biggr]. \label{gpr:2.5}
\end{equation}
in terms of the flavor non-singlet quark distributions, valid for large $x$.
We can then define the interative approximation:
\begin{equation}
A_G^{n+1}=\Biggl[{{\Delta P_{Gq} \otimes \Delta q+\Delta P_{GG}\otimes
(A_G^n\cdot G)} \over {P_{Gq} \otimes q+P_{GG}\otimes G}}\Biggr],
\label{gpr:2.6}
\end{equation}
which should converge for large enough n. It is important to note that
(\ref{gpr:2.6}) determines the form of $A_G(x)$ from the three distributions,
$\Delta q(x,t)$, $q(x,t)$ and $G(x,t)$, extracted from data. The spin-weighted
gluon asymmetry is then determined explicitly by $\Delta G(x,t)=A_G(x)\cdot
G(x,t)$. 

At small-$x$, we can also argue that ${{\partial A_G}\over {\partial t}}
=0$. We can parametrize the asymmetry in the form $A_G(x,t)=A_G(x)+
\epsilon(x,t)$, where $\epsilon(x,t)$ is a correction term which necessarily
vanishes at large $t$. Then,
\begin{equation}
{{\partial A_G(x,t)}\over {\partial t}}={{\partial \epsilon(x,t)}\over
{\partial t}}={1\over G}\Biggl[{{\partial \Delta G}\over {\partial t}}
-A_G(x,t){{\partial G}\over {\partial t}}\Biggr]. \label{gpr:2.7}
\end{equation}
Now, insert the expression for $A_G(x,t)$ in terms of $A_G(x)$ and $\epsilon$
into (\ref{gpr:2.7}) to get
\begin{equation}
{{\partial \epsilon(x,t)}\over {\partial t}}=-{{\epsilon(x,t)}\over
{G(x,t)}}\cdot{{\partial G(x,t)}\over {\partial t}}. \label{gpr:2.8}
\end{equation}
This implies that $\epsilon(x,t)\cdot G(x,t)$ is scale invariant. Thus, at
small-$x$, the growth in $G$ predicted by the evolution equations ensures that
$\epsilon\to 0$ at large $t$ and that $A_G$ maintains its $x$-dependent shape
asymptotically in $t$. 
 
For the starting distributions in eq. (\ref{gpr:2.5}) and the iterations of
eq. (\ref{gpr:2.6}), we use the polarized quark distributions outlined by 
GGR\cite{ggr} and the CTEQ4M unpolarized distributions.\cite{cteq} The
evolution was performed in LO, since the NLO contributions to the splitting
kernels, calculated in ref.\cite{werner}, are most dominant at small-$x$,
where the asymmetry is the smallest. Work is in progress to ensure that the
effects of NLO are not significant for the ratio $\frac{\Delta G}{G}$.
The iteration is relatively stable and converges within a couple of cycles.
Some models of $G(x)$ converge more uniformly than others. The resulting shape
of $A_G(x)$ is shown in Figure 1. This shape implies a larger polarized gluon
distribution than the $xG$ model of GGRA\cite{ggr}, so spin asymmetries which 
depend upon $\Delta G$ are enhanced. 

\begin{figure}
\begin{center}
\input{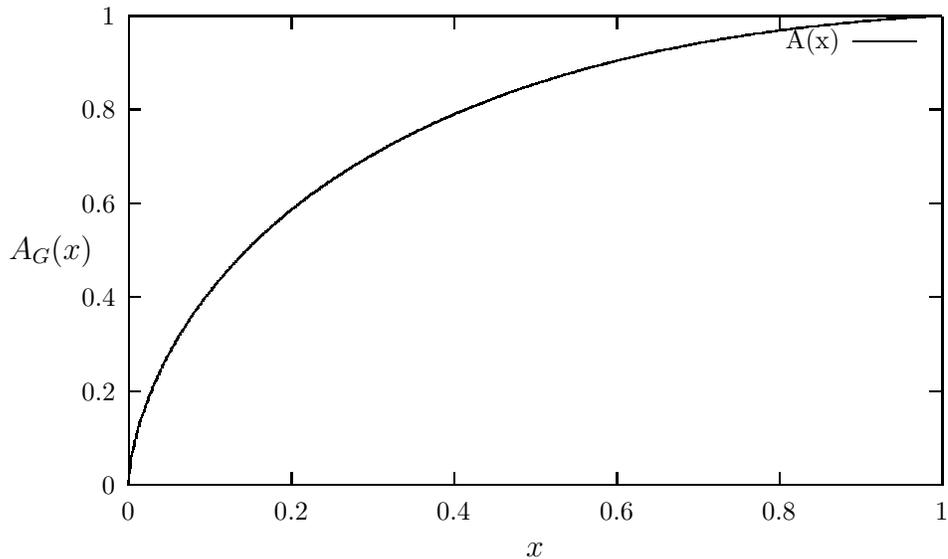}
\end{center}
\caption{The gluon asymmetry $\frac{\Delta G}{G}$ plotted as a function of
$x$}
\end{figure}

\section{Experimental Tests of $A_G(x)$ and $\Delta G$}

There are a number of existing and planned experiments are suitable for
measuring either $A_G(x)$ or a combination of $\Delta G(x,Q^2)$ and $G(x,Q^2)$. 
The HERMES experimental group at DESY has measured the longitudinal cross
section asymmetry $A_{\|}$ in high-$p_T$ hadronic photoproduction.\cite{hermes}
From this and known values of $\frac{\Delta q}{q}$ from DIS, a value for
$A_G(x_G)$ can be extracted. Here, $x_G=\hat{s}/2M\nu$ is the nucleon momentum
fraction carried by the gluon. Our corresponding value at $x_G=0.17$ is within
one $\sigma$ of the quoted value of $A_G=0.41\pm 0.18$ (stat.) $\pm 0.03$
(syst.).   

Both direct-$\gamma$ production and jet production at RHIC provide the best
means of extracting information about $\Delta G(x,Q^2)$ and $G(x,Q^2)$ 
separately.\cite{rhic,gr} The kinematic regions of STAR and
PHENIX can determine $A_G$ over a suitable range of $x_{Bj}$ to test this
model of the gluon asymmetry. Coupled with additional direct measurements of
$A_G(x_G)$ from HERMES, an appropriate cross check of $\Delta G(x)$ and $G(x)$
can be made. Since this model of $\frac{\Delta G}{G}$ implies a larger 
polarized glue than the GGRA model used in ref.\cite{gr}, all of the asymmetries
for direct-$\gamma$ and jet production should be enhanced, making them
easier to distinguish from the other parmetrizations of $\Delta G$.

The COMPASS group at CERN.\cite{compass} plans to extract $A_G$ from the photon
nucleon asymmetry, $A_{\gamma N}^{c\bar{c}}(x_G)$ in open charm muo-production,
which is dominated by the photon-gluon fusion process. This experiment should
be able to cover a wide kinematic range of $x_G$ as a further check of this
model. The combination of these experiments will be a good test of the
assumptions of our gluon asymmetry model and a consistency check on our
knowledge of the gluon distribution in the nucleon and its polarization.

\end{document}